\documentstyle[12pt]{article}

\advance\textwidth 2.0cm
\advance\oddsidemargin -1.0cm

\begin{document}

\title{
Curvature-based gauge-invariant\\ perturbation theory for gravity:\\
a new paradigm
}

% repeat the \author\address pair as needed
\author{Arlen Anderson, Andrew M.~Abrahams\cite{AMA}, and 
Chris Lea \\
Department of Physics and Astronomy\\
         University of North Carolina\\
         Chapel Hill, NC 27599-3255
          }

%\date{\today}
\date{Jan. 19, 1998}

\maketitle
\vspace{-11cm}
\hfill IFP-UNC-522

%\hfill TAR-UNC-05*

\hfill gr-qc/9801071
\vspace{9cm}

\begin{abstract}
A new approach to gravitational gauge-invariant perturbation theory
begins from the fourth-order Einstein-Ricci system, a hyperbolic
formulation of gravity for arbitrary lapse and shift whose 
centerpiece is a wave equation for curvature.  In the Minkowski
and Schwarzschild backgrounds, an intertwining
operator procedure is used to separate 
physical gauge-invariant curvature perturbations from unphysical ones. 
In the Schwarzschild case, physical variables are found which 
satisfy the Regge-Wheeler equation in both odd {\it and even} parity.
In both cases,
the unphysical "gauge'' degrees of freedom are identified with violations 
of the linearized Hamiltonian and momentum constraints, and they are found 
to evolve among themselves as a closed subsystem.  If the constraints are 
violated, say by numerical finite-differencing, this system
describes the hyperbolic evolution of the constraint violation.  
It is argued that an underlying {\it raison d'\^etre} of causal hyperbolic
formulations is to make the evolution of constraint violations 
well-posed. 
\end{abstract}

\newpage

\section{Introduction}
\label{sec:intro}

The paradigmatic application of perturbation theory in general
relativity is to describe distortions of the Schwarzschild black 
hole\cite{Cha83}. Moncrief's
classic gauge-invariant treatment\cite{Mon74} of this problem
is a remarkable piece of work, but to many its success has an air of
the magical to it.  The principal motivation for this paper is to use
the Schwarzschild problem to introduce a powerful and elegant new
method for gravitational gauge-invariant perturbation theory.
This method is practically algorithmic and follows
naturally from a hyperbolic formulation
of the Einstein equations\cite{aacby95a,aacby95b,aacby96a}, referred
to hereafter as the Einstein-Ricci formulation, whose centerpiece is a wave
equation for curvature.  Other applications of this new hyperbolic 
formulation have been pursued in \cite{YCB96a,YCB96b}. 

Besides serving as an example of a general technique, 
several new insights into
black hole perturbation theory arise: 
The first of these is the recognition that the
natural gauge-invariant perturbations
are curvature perturbations, not metric perturbations, though
one can get ``back to the metric'' if one desires.
Many people have shared the intuition that Einstein's
theory should truly be a theory of propagating curvature, but
here that ideal is explicitly realized.  Perhaps more
surprising is the discovery that the linearized Hamiltonian
and momentum constraints themselves constitute the non-physical
gauge degrees of freedom.  Violations of the constraints 
represent perturbations that take the background solution away
from the constraint hypersurface defining the physical theory.
Finally, the physical gauge-invariant variables satisfy
Regge-Wheeler equations\cite{ReW57}, coupled to constraint violations,
in both odd {\it and even} parity.

The following schematic form is found in detail below for the
system of equations describing the evolution of the
radial part of perturbations about the Schwarzschild background
$$
ds^2 = -N^2 dt^2 + N^{-2} dr^2 + r^2 d\theta^2 + r^2 \sin^2 \theta\phi^2,
$$
$N^2 = 1 - 2 M/r$, on a $t=$const., $K_{ij}=0$ slice.  
The gauge-invariant physical 
degrees of freedom are $A_u$ in odd-parity and $A_g$ in even-parity. (Their
expression in terms of the radial parts of perturbations of 3+1 variables 
will be given below.) 
Momentum constraint violations are encoded in $c_{u\theta}$ for odd-parity
and $c_{g\theta}$ and $c_{gr}$ in even-parity.  Violation of the
Hamiltonian constraint is encoded in $c_h$.  These equations
are valid in a neighborhood of the Schwarzschild initial data,
which is a ``point'' on the physical constraint hypersurface in the
phase space of Einsteinian initial data.  They 
reveal how perturbations of Schwarzschild behave even when
the constraints are violated perturbatively.  The equations are 
\begin{equation}
\label{Au_schem}
\biggl( -{\partial}_t^2 + (N^2 {\partial}_r)^2 - V_{RW} \biggr) A_u
- V_{u\theta} c_{u\theta} = 0
\end{equation}
\begin{equation}
\label{Ag_schem}
\biggl( -{\partial}_t^2 + (N^2 {\partial}_r)^2 - V_{RW} \biggr) A_g
- V_{gr} c_{gr} -  V_{g\theta} c_{g\theta} - V_{gh} c_h = 0
\end{equation}
\begin{equation}
\label{momu_schem}
\biggl( -{\partial}_t^2 + (N^2 {\partial}_r)^2 - V_{u\theta\theta} \biggr) 
c_{u\theta} = 0
\end{equation}
\begin{equation}
\label{momg_schem}
\hspace{-0.5cm}\left( \begin{array}{ccc}
-{\partial}_t^2 + {\bar\nabla}^k {\bar\nabla}_k - V_{hh}&-V_{hr}& 
- V_{h\theta} \\[0.4cm]
-V_{rh} & -{\partial}_t^2 + (N^2 {\partial}_r)^2 - V_{rr} & 
- V_{r\theta} \\[0.4cm]
-V_{\theta h} & - V_{\theta r} & -{\partial}_t^2 + (N^2 {\partial}_r)^2 
- V_{g\theta\theta}
\end{array} \right)
\left( \begin{array}{c}
c_h \\[0.4cm]
c_{gr} \\[0.4cm]
c_{g\theta}
\end{array} \right)
= 0,
\end{equation}
where $V_{RW}$ is the Regge-Wheeler potential\cite{ReW57}.

There are two immediate significant observations to make about this system of
equations.  The constraint variables evolve among themselves under 
the hyperbolic sub-system of equations (\ref{momu_schem}), 
(\ref{momg_schem}).  If 
they and their time derivatives vanish initially,  they are guaranteed 
to vanish for all time.  This is conjectured to be a general feature of
constrained hyperbolic systems and is discussed further below.  Next, when
the constraints are satisfied, so the constraint variables vanish,
both the odd- and even-parity physical perturbations satisfy
the Regge-Wheeler equation\cite{ReW57}. 
This somewhat surprising result runs
counter to folklore which attributes the existence of the
familiar Zerilli equation\cite{Zer70} for even-parity perturbations somehow
to parity.
The new derivation reveals that parity is not the issue. 

Chandrasekhar\cite{Cha83} has observed that because there is no difference
between even and odd parity in the Newman-Penrose formalism, there
is no reason to expect different equations for the different
parities.  This was part of the motivation that led him to construct 
the transformations\cite{Cha75} between the Zerilli and Regge-Wheeler 
equations and their Newman-Penrose analog, the Bardeen-Press 
equation\cite{BaP72}, that make their isospectrality clear.   Such isospectral
transformations between equations are ubiquitous\cite{And,And91}, but 
transformations
which relate similarly simple potentials arise only under
special conditions (cf. e.g. \cite{AnP}), something which could well be
attributed to accident.  It is clear from the present work that the
existence of the isospectral transformation allows one to reach
both the Regge-Wheeler and Zerilli equations, in either parity, but 
that the Regge-Wheeler
equation is preferable on the grounds of simplicity.  In particular,
the Regge-Wheeler equation has regular singular points only
at physically significant locations while the Zerilli equation has a further
regular singular point at an angular momentum dependent location.

An intuitive argument helps to explicate the structure of the 
evolution equations (\ref{Au_schem})-(\ref{momg_schem}).  In gauge-invariant
perturbation theory, the natural candidate variables for perturbation
are those which vanish in the background\cite{ElB}.  In a 
constrained theory, the constraints themselves 
furnish a natural subset of such variables:  The constraints are constructed 
from the degrees of freedom of the full unconstrained theory; they
are satisfied, i.e. vanish, in the background; under arbitrary perturbations, 
they are generally nonzero.  Indeed, just as satisfaction of
the constraints defines the physical sector of the theory, their
violation (partially) parametrizes the theory away from the physical sector
of the theory.  The variables based on the constraints will be
referred to as constraint variables.  In distinction, the variables
which parametrize the constraint hypersurface will be referred to
as physical variables.  The constraint variables are a wise choice to 
describe some of
the ``gauge'' degrees of freedom of a constrained theory, gauge in the 
sense of characterizing unphysical aspects of motion.   

By a constrained {\it hyperbolic} theory, we mean a constrained theory which
is well-posed, and in particular one in which the constraints are 
guaranteed to remain satisfied provided they are satisfied initially.
It seems intuitively clear that if such a theory were expressed 
in terms of constraint variables and physical variables, 
the system of equations would split much as 
(\ref{Au_schem})-(\ref{momg_schem}) do.
There would be a subsystem in which the constraint variables evolve
among themselves, and there would be further equations in which the
physical variables couple to the constraint variables.  Physical
variables could appear in the constraint subsystem only nonlinearly,
if at all, multiplied by constraint variables, or else their nonzero 
presence would act as a source to drive the constraint variables away 
from zero, given vanishing initial data.  Because of this, the 
physical variables cannot appear 
in the constraint variable subsystem in the perturbative setting.   
That a system of equations would admit such a closed internal
system of equations is obviously special, but it is the feature which
distinguishes a constrained hyperbolic system from an unconstrained
one.   This qualitative analysis is expected to hold in 
the fully nonlinear theory, and work is in progress to demonstrate 
this\cite{AnY}. 

One of the subtle issues that arises in constrained physical theories 
like general relativity is that while the theory is hyperbolic on 
physical grounds, its mathematical representation in redundant variables 
may not be.  In other words, were the theory reduced to the
true degrees of freedom, it would be a hyperbolic theory.  However, when
expressed in redundant variables, the theory may not impose hyperbolic 
evolution on unphysical combinations of variables, 
{\it e.g.} the constraint variables.  
Einstein's equations, viewed as a system of differential equations for
the metric, are an example of a constrained system which is physically
hyperbolic but not mathematically so.  If further restrictions are
applied, e.g. through special coordinate conditions, a modified system
of equations can be made hyperbolic\cite{cby,FiM}.  

The several new
recent hyperbolic formulations follow a different route to 
hyperbol\-icity\cite{aacby95a,aacby95b,aacby96a,hyper}. 
By various extensions, the redundant sector of the theory is enlarged, so 
the whole theory becomes manifestly hyperbolic.  A key virtue
of achieving a hyperbolic formulation is that the system of equations
is well-posed, so that, in particular in causal hyperbolic formulations, this
guarantees that violation of the constraints evolves in a 
predictable, though not necessarily stable, fashion\cite{AnY}.  
An ill-posed theory is vulnerable to catastrophic breakdown: 
arbitrarily small perturbations may lead to arbitrarily large
deviations arbitrarily quickly.  In a well-posed theory, there
may be exponentially growing modes, which may in turn be
identified as instabilities, but their growth rate is bounded. 

In numerical simulations, the urgency of these considerations comes
to the fore.  Except in special cases, the very implementation of
the constrained system will introduce constraint violation because
finite-differencing generally does not respect the constraints.
While there exist numerical methods for evolving ill-posed systems,
there are many more highly developed methods for handling hyperbolic
systems.  One can choose to re-solve the constraints at each time
step to try to remain near the constraint hypersurface, or one
can face the full theory and address the issue of controlling
constraint violation directly.  In the perturbative example here,
once the splitting into constraint and physical variables has been
made, preservation of the constraints under differencing is no longer
an issue---this reflects the fact that the theory has been reduced
to the linearized true degrees of freedom.   On the other hand, one can 
undertake a stability analysis of the system with constraint violations
which will reflect on the theory in the original variables, and 
thereby gain insight into the nature of instabilities likely to
appear in a numerical simulation of the full theory in the original variables. 
Work on these issues is in progress.

\section{Intertwining procedure}

We outline the procedure for gravitational gauge-invariant perturbation
theory as follows: Consider perturbations about the Schwarzschild
background. The fourth-order form of the Einstein-Ricci
formulation\cite{aacby96a} is a hyperbolic system of a type termed
``hyperbolic non-strict''\cite{LeO}. One equation of this system is a wave
equation for the time-derivative of the extrinsic curvature,
${\hat\partial_0} K_{ij}$ (${\hat\partial_0}
\equiv {\partial}/{\partial} t - {\cal L}_\beta$, where ${\cal L}_\beta$ is
the Lie derivative along the shift $\mbox{\boldmath $\beta$}$). Because
${\hat\partial_0} K_{ij}$ is part of the decomposition of the Riemann
tensor $R_{0i0j}$ in 3+1 variables, this wave equation propagates
curvature. The time derivatives of the components of the extrinsic
curvature vanish in the Schwarzschild background, so their perturbations
are necessarily gauge-invariant. They are chosen as the perturbative
quantities and will be referred to hereafter as curvature perturbations.
Their selection accords with the principle that the natural candidates for
gauge-invariant quantities are those which vanish in the
background\cite{ElB}. Note, in contrast, that the Riemann tensor itself
$R_{0i0j}$ does not vanish in the background. The Einstein-Ricci system is
hyperbolic for arbitrarily specified lapse and shift, which consequently do
not have to be perturbed.

There are six components of ${\hat\partial_0} K_{ij}$, and one has
six coupled wave equations in terms of them.  In addition, there
are four ``constraints,'' formed from time-derivatives
of the three momentum constraints and the Hamiltonian constraint.
This leaves two independent equations to be found.  If one makes
a tensor spherical harmonic decomposition of the perturbations of 
${\hat\partial_0} K_{ij}$,
the latter naturally divide into odd and even parity.  The 
two dynamical equations and one constraint with odd parity 
lead to the Regge-Wheeler equation.  This leaves four dynamical
equations and three constraints with even parity.

The isolation of a wave equation from the even
parity system is the technical crux of the gravitational perturbation 
calculation.
An analogy will make the procedure clear.  Suppose that one 
had four linear algebraic equations in four unknowns.  To solve them,
one would take linear combinations of the equations to isolate
linear combinations of the unknowns. Equivalently,
one diagonalizes the matrix of coefficients in the equations by
transforming to a different basis.

In the present case, we have four linear differential equations in four
unknowns.  We construct linear differential combinations of the equations
to isolate linear differential combinations of the unknowns.
We do not diagonalize the matrix of differential operators,
but we re-group the variables to split the differential equations
into partially uncoupled form which is diagonal when 
the constraints hold.
We do this by grouping the original variables into combinations
which constitute the constraints and new variables which are
independent of the constraints.  
This is accomplished by an intertwining transformation\cite{And,And91} 
in which one matrix of
differential operators $M_1$ is transformed into another $M_2$ by a matrix
operator ${\cal D}$
\begin{equation}
\label{ioDef}
M_2 {\cal D} = {\cal D} M_1.
\end{equation}
If ${\cal D}$ were invertible, one would have the familiar expression
${\cal D} M_1 {\cal D}^{-1} = M_2$ for a basis change.
When the constraints hold, the matrix $M_2$ is diagonal;
when they do not, the matrix $M_2$ is simply a matrix
in a different basis.

Nonperturbatively, the momentum constraints are 
\begin{equation}
R_{0j} = -N \biggl( {\bar\nabla}^k K_{jk} - {\bar\nabla}_j H \biggr)=0
\end{equation}
where $H=K^k\mathstrut_k$ is the trace of the extrinsic curvature
and overbars indicate spatial quantities, here spatial covariant
derivatives.
The perturbed time-derivative of the momentum constraints are 
differential linear combinations of the curvature perturbations
${\hat\partial_0} K_{ij}$.
They are thus appropriate combinations to use in re-organizing
the system of four linear differential equations.  As displayed in
(\ref{momu_schem}), (\ref{momg_schem}), the momentum constraint variables 
evolve
among themselves together with the Hamiltonian constraint variable. 

The Hamiltonian constraint itself 
$$G^0\mathstrut_0= - {1\over 2}(H^2 - K^{ij} K_{ij} + {\bar R})=0$$ 
involves only metric perturbations when perturbed about a
${\underline{K}}_{ij}=0$ slice (where the underline indicates a background
quantity). As a spatial constraint, this constraint may be violated when
evaluated on a general metric and extrinsic curvature. This reflects
violation of the Hamiltonian constraint. Taking two time derivatives
converts the metric perturbations to curvature perturbations and gives an
equation for the evolution of violation of the Hamiltonian constraint. This
is simply the time derivative of the doubly-contracted Bianchi identity
$\nabla^\mu G_{\mu 0}=0$, which can therefore be read as an equation
evolving violations of the Hamiltonian constraint through coupling
to violations of the momentum constraints (cf. \cite{Fri97}).

Separately, the relation
$$R^0\mathstrut_0=-N^{-1} {\hat\partial_0} H + K_{ij} K^{ij} 
- N^{-1} {\bar\nabla}^k {\bar\nabla}_k N =0$$
stands as a dynamical ``constraint'' coupling curvature and metric
perturbations in perturbation theory. This equation is not
part of the Einstein-Ricci system and hence may be violated.
Two time derivatives again convert the metric perturbations to curvature 
perturbations. 

The identity
\begin{equation}
\label{GRid}
R^k\mathstrut_k = -2 G^0\mathstrut_0 + R^0\mathstrut_0
\end{equation}
relates the two constraints to the field equations.  The second 
time derivative of $R^k\mathstrut_k$ occurs as part of
the dynamical equations which define the fourth-order Einstein-Ricci
theory and is therefore part of an equation which does not admit 
violations.  Thus, two time derivatives of (\ref{GRid}) establishes the 
correlation between violations of $G^0\mathstrut_0=0$  and of
$R^0\mathstrut_0=0$.  In view of this correlation, through a useful 
abuse of language,  $R^0\mathstrut_0$ will be referred to also as 
the Hamiltonian constraint, and the so-called Hamiltonian constraint variable
$c_h$ in (\ref{Au_schem})-(\ref{momg_schem}) reflects violations of the 
$R^0\mathstrut_0$ constraint. Violations of $G^0\mathstrut_0$ can be eliminated
from the second time derivative of (\ref{GRid}) using the doubly-contracted 
Bianchi identity 
mentioned above, resulting in the wave equation in (\ref{momg_schem}) 
for violations of $R^0\mathstrut_0=0$.  Details of this will be given below.  

Thus, three of the four equations have been ``diagonalized'' already, and
it is not necessary to begin from a general matrix ansatz for
the intertwining matrix ${\cal D}$, though this could have been done.  Let
$g_k(r,t)$, $k=1,\ldots,4$, be the radial parts of the
even-parity curvature perturbations, and let ${\hat L}_k(g_1,\ldots,g_4;r,t)=0$,
$k=1,\ldots,4$, be the associated dynamical coupled second-order
differential equations for them.  (Note that ${\hat L}_k$ is not simply the
differential operator, but the full equation; 
this is indicated notationally by the hat.)
It is sufficient to consider a first-order differential linear combination
of the four dynamical equations to achieve a single wave
operator acting on the same differential linear combination
of radial curvature perturbations
\begin{equation}
\label{LitoRW}
\sum_{k=1}^4 (b_k {\partial}_r + a_k) {\hat L}_k = L_{RW} \sum_{k=1}^4
(b_k {\partial}_r + a_k) g_k + {\hat f}(R_{00},R_{0j}),
\end{equation}
where
\begin{equation}
\label{RWOp}
L_{RW} = - {\partial}_t^2 + (1-{2M\over r}) {\partial}_r (1-{2M\over r}) 
{\partial}_r - V_{RW}(r)
\end{equation}
is a wave operator of Regge-Wheeler form and ${\hat f}$ is a linear 
differential operator applied to the constraints.  The potential
\begin{equation}
\label{RWpot}
V_{RW}(r)=(1- {2M\over r}) ({\ell(\ell+1) \over r^2} - {6 M\over r^3})
\end{equation}
is the Regge-Wheeler potential---it could be left undetermined initially and
solved for self-consistently.  (Alternatively, one could attempt to reach
the Zerilli potential starting from a second-order differential
combination.)
The $a_i$ and $b_i$ are functions of $r$ to be determined by
equating like coefficients of derivatives of the $g_k$'s term-by-term.

A simplification is achieved
by recognizing that the momentum constraints can be used to
eliminate linear spatial derivatives of particular radial coefficients in
favor of other terms. This allows one to set two of the $b_k$
to zero without loss of generality.
The equations for the $a_k$ and $b_k$ obtained from (\ref{LitoRW})
by equating terms are over-determined.  After determining $a_k$ and
$b_k$ from a
sufficient set of equations, the remaining equations are consistency
conditions.  If the equations are not consistent, an intertwining
of the chosen form does not exist, though one of a different form,
e.g. higher order differential or infinite order by means of integral
operators, may exist.

The operator transformation between the Regge-Wheeler and Zerilli equations
that Chandrasekhar\cite{Cha83} discusses is also an example of 
intertwining\cite{AnP,Whi}.
Because intertwining transformations compose, when
diagonalizing the system above, either equation can be reached by
using an appropriate linear differential combination of the curvature
perturbations.
One has
\begin{eqnarray}
\label{RWtoZ}
D_1 L_{RW} &=& L_Z D_1, \\
D_2 L_Z &=& L_{RW} D_2, \nonumber
\end{eqnarray}
where the Regge-Wheeler operator $L_{RW}$ is given by (\ref{RWOp}) and
the Zerilli operator is
\begin{equation}
\label{ZerOp}
L_Z =  - {\partial}_t^2 + (1-{2M\over r}) {\partial}_r (1-{2M\over r}) 
{\partial}_r
-V_Z(r)
\end{equation}
with the Zerilli potential\cite{Cha83,Zer70}
\begin{equation}
V_Z(r) = {2 N^2 \over r^3 ( n r + 3 M)^2} \biggl((n+1)n^2 r^3 + 3 M n^2 r^2 +
9  M^2 n r + 9 M^3\biggr),
\end{equation}
where $N^2= 1 - 2M/r$ and $n=(\ell-1)(\ell+2)/2$.
It is important to note that one can pass in either direction by
means of differential operators; the inverse transformation need
not be an integral operator.  The intertwining operators are\cite{AnP}
\begin{eqnarray}
\label{ioRWZer}
D_1 &=& (1-{2 M\over r}) {\partial}_r  + {3M(r - 2 M)\over r^2 (nr + 3M)} 
+\omega, \\
D_2 &=& (1-{2 M\over r}) {\partial}_r  - {3M(r - 2 M)\over r^2 (nr + 3M)}
-\omega, \nonumber
\end{eqnarray}
where $\omega=n(n+1)/(3M)=(\ell-1)\ell(\ell+1)(\ell+2)/(12 M)$.
The intertwining relations (\ref{RWtoZ}) become more transparent
when one recognizes that they are a consequence of associativity,
$D_1 (D_2 D_1) = (D_1 D_2) D_1$ and $D_2 (D_1 D_2) = (D_2 D_1) D_2$,
because the radial parts of the Regge-Wheeler and Zerilli operators factorize
\begin{eqnarray}
L_{RW} &=& D_2 D_1 -{\partial}_t^2 + \omega^2, \\
L_Z &=& D_1 D_2 -{\partial}_t^2 + \omega^2.\nonumber
\end{eqnarray}
This factorization property is very suggestive, but it turns out not
to be the best property to generalize intertwining.

\section{Einstein-Ricci formulation}

As shown in~\cite{aacby95a,aacby95b,aacby96a}, the dynamical part of 
Einstein's equations can be cast in hyperbolic form in 3+1 language 
as the Einstein-Ricci formulation.  Consider a globally 
hyperbolic manifold of topology $\Sigma\times R$ with the metric
\begin{equation}
ds^2 = -N^2 dt^2 +g_{ij} (dx^i +\beta^i dt) (dx^j +\beta^j dt),
\end{equation}
where $N$ is the lapse, $\beta^i$ is the shift, and $g_{ij}$ is the 
spatial metric.  Introduce the non-coordinate co-frame,
\begin{equation}
\label{nocoord}
\theta^0 = dt, \quad
\theta^i = dx^i+\beta^i dt.
\end{equation}
with corresponding dual (convective) derivatives
\begin{equation}
\label{nocodual}
\partial_0 = \partial/\partial t -\beta^i \partial/\partial x^i, \quad
\partial_i = \partial/\partial x^i.
\end{equation}
 
The natural time derivative for evolution is\cite{Yor79}
\begin{equation}
\hat\partial_0=\partial_0 +\beta^k \partial_k -{\cal L}_{\beta}
=\partial/\partial t -{\cal L}_{\beta},
\end{equation}
where ${\cal L}_{\beta}$ is the Lie derivative in a time slice $\Sigma$ along
the shift vector.  In combination with the lapse as
$N^{-1}\hat\partial_0$, this is the derivative with respect to proper time
along the normal to $\Sigma$.
The extrinsic curvature $K_{ij}$ of
$\Sigma$ is defined as
\begin{equation}
\label{Kij}
\hat\partial_0 g_{ij}= -2 N K_{ij}.
\end{equation}
{}From this follows the equation for the evolution of the Christoffel
connection
\begin{equation}
\label{Gam}
{\hat\partial_0} {\bar\Gamma}^k\mathstrut_{ij} = 
-g^{mk}[{\bar\nabla}_j (N K_{im}) 
+{\bar\nabla}_i(N K_{mj}) - {\bar\nabla}_m(NK_{ij}) ].
\end{equation}
Barred quantities are three-dimensional.

The dynamical Einstein equations, $R_{ij}= \rho_{ij}$, where $\rho_{ij}$
is a matter source, are equivalent to a {\it third-order Einstein-Ricci 
system}.  The system is labelled ``third-order'' by the equivalent of
the highest number of derivatives of $g_{ij}$ that can occur in the theory.  
(These derivatives need not and generally do not appear explicitly.)
The third-order Einstein-Ricci system is
obtained from (\ref{Kij}) and a wave equation for $K_{ij}$
\begin{equation}
\label{weq}
N^2 \hat{\mbox{\kern-.0em\lower.3ex\hbox{$\Box$}}} K_{ij} + N J_{ij} + N S_{ij} 
= N \Omega_{ij} \equiv N({\hat\partial_0} R_{ij}
-{\bar\nabla}_i R_{0j} - {\bar\nabla}_j R_{0i}),
\end{equation}
where
\begin{equation}
\hat{\mbox{\kern-.0em\lower.3ex\hbox{$\Box$}}} =
-N^{-1} {\hat\partial_0} N^{-1} {\hat\partial_0} 
+ {\bar\nabla}^k {\bar\nabla}_k.
\end{equation}
Here, $\Omega_{ij}$ is a matter source term which, as a consequence of 
the field equations, vanishes in vacuum.
For this paper, we restrict attention to vacuum spacetimes.
$J_{ij}$ is a nonlinear self-interaction term.
If we denote the trace of the extrinsic curvature by
$H=K^{k}\vphantom{|}_{k}$, $J_{ij}$ is given by
\begin{eqnarray}
\label{Jij}
J_{ij}&=& \hat\partial_0 (H K_{ij} - 2 K_{i}\vphantom{|}^{k} K_{jk})
 +(N^{-2}\hat\partial_0 N+ H)\bar\nabla_i \bar\nabla_j N \nonumber \\
&&\hspace{-0.75cm}
-2N^{-1}(\bar\nabla_k N) \bar\nabla_{(i}(N K^{k}\vphantom{|}_{j)})
+3 (\bar\nabla^k N) \bar\nabla_k K_{ij} \\
&&\hspace{-0.75cm} +N^{-1}K_{ij} \bar\nabla^k (N\bar\nabla_k N)
-2 \bar\nabla_{(i}(K_{j)}\vphantom{|}^{k}\bar\nabla_k N)
+N^{-1} H\bar\nabla_i\bar\nabla_j N^2
\nonumber \\
&&\hspace{-0.75cm}
+2 N^{-1} (\bar\nabla_{(i} H)(\bar\nabla_{j)} N^2)
 -2N K^{k}\vphantom{|}_{(i}\bar R_{j)k}
-2N \bar R_{kijm}K^{km}, \nonumber
\end{eqnarray}
where $M_{(ij)}={1\over 2} (M_{ij}+M_{ji})$.  
The three-curvatures can be expressed in terms of
four-curvatures and then eliminated using the
field equations. 
 
Finally, $S_{ij}$ is a slicing term,
\begin{equation}
S_{ij}=-N^{-1}\bar\nabla_i \bar\nabla_j(\hat\partial_0 N +N^2 H).
\end{equation}
This must be equal to a functional involving fewer than
third derivatives of the metric to assure the hyperbolic
(wave) nature of the equation (\ref{weq}). 
Two simple ways to do this are the following.
One may impose the harmonic slicing condition 
\begin{equation}
\label{d0N}
\hat\partial_0 N +N^2 H=0,
\end{equation}
in which case
\begin{equation}
\label{Sharm}
S_{ij}=0.
\end{equation}
Alternatively, one may specify the mean curvature by demanding $H=h(x,t)$. 
The lapse function $N$ is then determined by solving the elliptic equation
\begin{equation}
\label{Nell}
\bar\nabla^k\bar\nabla_k N= -\hat{\partial}_0 h(x,t) + N K_{ij} K^{ij}.
\end{equation}
The special case of maximal slicing, $H\equiv 0$, gives
\begin{equation}
\label{Smax}
S_{ij}=-N^{-1}\bar\nabla_i \bar\nabla_j \hat\partial_0 N.
\end{equation}
In this case, the Einstein-Ricci system is mixed 
hyperbolic-elliptic\cite{cby96}.
 
With harmonic slicing, the dynamical part of Einstein's equations are
given by the definition of the extrinsic curvature (\ref{Kij}), the wave
equation (\ref{weq}), and the harmonic slicing condition (\ref{d0N}).  (In 
this case, all the equations of motion are equivalent to a first-order 
symmetric hyperbolic system\cite{aacby95a,aacby95b} with characteristics 
the light cone and the direction orthogonal to the time slices.)
With slicing given by specified mean curvature, the
dynamical equations are (\ref{Kij}) and (\ref{weq}), and
the lapse is determined by interleaving the solution of
the elliptic equation (\ref{Nell}) on each time-slice.

To complete the new formulation of the Einstein equations, we
specify the initial Cauchy data.  Initial $g_{ij}$ and $K_{ij}$
must be chosen compatible with the Gauss-Codazzi (Hamiltonian and 
momentum) constraints.  The Hamiltonian constraint is
\begin{equation}
G^{0}\mathstrut_0={1\over 2}(R^0\mathstrut_0 - R^k\mathstrut_k) =
-{1\over 2}(\bar R +H^2 -K_{mk}K^{mk})=0,
\end{equation}
and the momentum constraints are
\begin{equation}
N G^{0}\mathstrut_{i}=\bar\nabla_k (K^{k}\mathstrut_{i} -
          \delta^{k}\mathstrut_{i} H)=0.
\end{equation}
These are treated as an elliptic system on the initial slice by the 
usual methods\cite{Yor79,cby80}.  Furthermore, to guarantee that the 
dynamical equations
produce a solution to Einstein's equations, it is also necessary
that $\hat{\partial}_0 K_{ij}$  be specified initially so that Einstein's
equations hold on the initial slice 
\begin{equation}
\label{ricci}
R_{ij}=\bar R_{ij} -N^{-1}\hat{\partial}_0 K_{ij}+H K_{ij} 
-2 K_{ik} K^{k}\mathstrut_j
-N^{-1}\bar\nabla_i \bar\nabla_j N,
\end{equation}
where $R_{ij}$ on the left hand side is replaced by its expression
in terms of the matter sources.  In the case of harmonic slicing, the
lapse is specified on the initial slice.  The shift, hidden in
${\hat\partial_0}$, is a freely specifiable function on spacetime and is
not a dynamical variable of the system. 

To obviate the need for special handling of the slicing term $S_{ij}$
and to allow the lapse to be specified freely,
one can take another time-derivative, apply a further constraint,
and pass to the fourth-order Einstein-Ricci formulation.  These equations 
proceed from (\ref{Kij}) and the equation
\begin{eqnarray}
\label{weq2}
N {\hat\partial_0}(N \hat{\mbox{\kern-.0em\lower.3ex\hbox{$\Box$}}} K_{ij}) 
+ N {\hat\partial_0}( J_{ij} +  S_{ij}) 
+ N {\bar\nabla}_i {\bar\nabla}_j ( N {\hat\partial_0} H -N^2 K_{mk} K^{mk} 
+ N {\bar\nabla}^k {\bar\nabla}_k N) =&& \nonumber \\
&&\hspace{-5in}= N \tilde \Omega_{ij} \equiv
N( {\hat\partial_0}^2 R_{ij} - 2 {\hat\partial_0} {\bar\nabla}_{(i} R_{j)0} 
+ {\bar\nabla}_i {\bar\nabla}_j R_{00}). 
\end{eqnarray}
The effect of including $R_{00}$ is to incorporate the Hamiltonian
constraint while cancelling the threatening highest derivative term
of $H$ in $S_{ij}$.  The Cauchy data are extended by specifying
${\hat\partial_0}^2 K_{ij}$, subject to the requirement (in vacuum) that 
${\hat\partial_0} R_{ij}=0$ hold on the initial slice.    
Both the lapse and the shift are
freely specifiable functions on spacetime and are not dynamical
variables.  The system (\ref{Kij}), (\ref{weq2}) is 
hyperbolic non-strict in the sense of Leray-Ohya\cite{LeO}.
In particular, it is well-posed with solutions in an appropriate
Gevrey (not Sobolev) class\cite{aacby96a}.

\section{Linearized gravity in the hyperbolic formulation}

Let us first consider a weak-field analysis of the equations above around a
flat spacetime background with a Minkowski metric. Let
$g_{ij}={\underline{g}}_{ij}+{g'}_{ij}$ and
$K_{ij}={\underline{K}}_{ij}+{K'}_{ij}$ where underlines denote background
values and primes the first order corrections. In the present case,
${\underline{g}}_{ij}=\delta_{ij}$ and ${\underline{K}}_{ij}=0$. In the
fourth-order Einstein-Ricci formulation, the lapse and shift can be
arbitrarily specified: allow a lapse perturbation $N={\underline{N}} + N'
=1 + N'$, but set the shift perturbation to zero, $\beta^i =
{\underline{\beta}}^i =0$. The use of a lapse perturbation is unnecessary,
but its inclusion will demonstrate this explicitly. (The lapse perturbation
$N'$ here should not be confused with $N'={\partial}_r N$ used later in the
paper.) At first order, (\ref{Kij}) becomes
\begin{equation}
\label{foKij}
{\partial}_t {g'}_{ij}=-2 {K'}_{ij}.
\end{equation}
The wave equation (\ref{weq2}) from the fourth-order Einstein-Ricci 
formulation is
\begin{equation}
\label{linweq2}
\hat{\mbox{\kern-.0em\lower.3ex\hbox{$\underline{\Box}$}}} 
{\partial}_t{K'}_{ij}
 -{\underline{\bar\nabla}}_i {\underline{\bar\nabla}}_j ({\partial}_t^2 N' 
 + {\partial}_t H') +
{\underline{\bar\nabla}}_i {\underline{\bar\nabla}}_j {R'}_{00} =0
\end{equation}
($\hat{\mbox{\kern-.0em\lower.3ex\hbox{$\underline{\Box}$}}} = 
-{\partial}_t^2 + {\underline{\bar\nabla}}^k {\underline{\bar\nabla}}_k$),
where ${\underline{\bar\nabla}}_i$ is the spatial covariant derivative with
respect to the background metric ${\underline{g}}_{ij}$. 

The weak-field form of the identity
\begin{equation}
\label{curv}
R_{0i0j}=N\hat{\partial}_0 K_{ij}+N^2 K_{ik} K^{k}\mathstrut_j
+N\bar\nabla_i \bar\nabla_j N
\end{equation}
is
\begin{equation}
\label{curvp}
{R'}_{0i0j} = {\partial}_t {K'}_{ij} 
+ {\underline{\bar\nabla}}_i {\underline{\bar\nabla}}_j N'.
\end{equation}
Its trace gives
\begin{equation}
{R'}_{00} = {\partial}_t H' 
+ {\underline{\bar\nabla}}^k {\underline{\bar\nabla}}_k N'.
\end{equation}
Using this in (\ref{linweq2}) gives
\begin{equation}
\label{linweq3}
\hat{\mbox{\kern-.0em\lower.3ex\hbox{$\underline{\Box}$}}} 
({\partial}_t {K'}_{ij} 
+ {\underline{\bar\nabla}}_i {\underline{\bar\nabla}}_j N') 
= \hat{\mbox{\kern-.0em\lower.3ex\hbox{$\underline{\Box}$}}} {R'}_{0i0j} =0 .
\end{equation}
This equation is clearly lapse independent.  Furthermore, the
perturbations are explicitly Riemann tensor perturbations.

There are six degrees of freedom to ${R'}_{0i0j}$, but we expect
only two physical gauge-invariant degrees of freedom.  The resolution
of this apparent paradox is to recognize that by tracing (\ref{linweq3})
one obtains an equation for ${R'}_{00}$
\begin{equation}
\hat{\mbox{\kern-.0em\lower.3ex\hbox{$\underline{\Box}$}}} {R'}_{00} =0,
\end{equation}
while taking a (spacetime) divergence and using the contracted Bianchi 
identity, 
\begin{equation}
\label{Bicon}
\nabla^i R_{0i0j} = -\nabla_0 R_{0j} + \nabla_j R_{00},
\end{equation}
one reaches an equation for ${\partial}_t {R'}_{0j}$
\begin{equation}
\hat{\mbox{\kern-.0em\lower.3ex\hbox{$\underline{\Box}$}}} 
{\partial}_t {R'}_{0j} =0.
\end{equation}
Thus, four of the six equations evolve constraints.  A general
perturbation may violate the constraints, and these four equations
determine the constraint violation evolution.  The physical degrees of
freedom are the remaining two degrees of freedom of ${R'}_{0i0j}$,
the transverse traceless parts.  

To make contact with traditional analyses, one can 
examine (\ref{curvp}) in more detail by reading it as an
equation for ${\partial}_t {K'}_{ij}$ and splitting ${K'}_{ij}$ into
a sum of transverse traceless, longitudinal traceless and trace parts
\begin{equation}
{K'}_{ij} = {K'}^{\rm TT}_{ij} + {K'}^{\rm LT}_{ij} + {1\over 3} g_{ij} H'.   
\end{equation}
The trace of (\ref{curvp}) is
\begin{equation}
\label{R00p}
{\partial}_t H' = {R'}_{00} 
- {\underline{\bar\nabla}}^k {\underline{\bar\nabla}}_k N'
\end{equation}
as above.  Note that ${R'}_{00}$ is not set to zero as 
traditionally done.  This is because a general perturbation
will violate the constraint, and it is easier to work with
free rather than constrained perturbations.
 
The longitudinal part can be identified by using the definition that the
divergence of the transverse part vanishes.  Split the extrinsic
curvature perturbation and the Riemann tensor into a sum of
transverse traceless and longitudinal (with trace) parts
\begin{equation}
{K'}_{ij} = {K'}^{\rm TT}_{ij} + {K'}^{\rm L}_{ij},\quad
{R'}_{0i0j} = {R'}^{\rm TT}_{0i0j} + {R'}^{\rm L}_{0i0j},
\end{equation} 
where the divergence of the TT parts is assumed to vanish.
The divergence of (\ref{curvp}) is then
\begin{equation}
\label{divcurvp}
{\partial}_t {\underline{\bar\nabla}}^i {K'}^{\rm L}_{ij} 
= {\underline{\bar\nabla}}^i {R'}^{\rm L}_{0i0j}
-{\underline{\bar\nabla}}^i {\underline{\bar\nabla}}_i 
{\underline{\bar\nabla}}_j N'.
\end{equation}
Using the definition of the perturbative momentum constraint
\begin{equation}
\label{R0jp}
{R'}_{0j} = - ({\underline{\bar\nabla}}^i {K'}_{ij} 
- {\underline{\bar\nabla}}_j H')
\end{equation}
and (\ref{R00p}) leads to the perturbative form of the contracted
Bianchi identity (\ref{Bicon})
\begin{equation}
-{\partial}_t {R'}_{0j} = {\underline{\bar\nabla}}^i {R'}^{\rm L}_{0i0j} 
-{\underline{\bar\nabla}}_j {R'}_{00}.
\end{equation}
This reveals the longitudinal nature of this identity.
Stripping the divergence away from (\ref{divcurvp}) and
removing the trace gives the tracefree longitudinal equation
\begin{equation}
{\partial}_t {K'}^{\rm LT}_{ij} = {R'}^{\rm L}_{0i0j} 
- {1\over 3} g_{ij} {R'}_{00}
-( {\underline{\bar\nabla}}_{(i} {\underline{\bar\nabla}}_{j)} N' 
- {1\over 3} g_{ij} {\underline{\bar\nabla}}^k {\underline{\bar\nabla}}_k
N').
\end{equation} 

Finally, the transverse traceless equation is
\begin{equation}
{\partial}_t {K'}^{\rm TT}_{ij} = {R'}^{\rm TT}_{0i0j}.
\end{equation}
If one further splits (\ref{foKij}) to obtain the transverse
traceless part, then one finds
\begin{equation}
-{1\over 2}{\partial}_t^2 {g'}^{\rm TT}_{ij} = {R'}^{\rm TT}_{0i0j},
\end{equation}
which is a well-known result.
One of the virtues of this analysis is that it clarifies the
role of the transverse and longitudinal parts of the Riemann
tensor and emphasizes that the true physical degrees of the freedom
of the linearized gravitational field lie in the transverse traceless part.

\section{Gauge-invariant perturbation theory: Schwarzschild}

The fourth-order Einstein-Ricci system is the natural one for
gauge-invariant perturbation theory.  Because the lapse and the
shift are freely specifiable, their perturbations do not need to
be considered.  Furthermore, because one is working with perturbations
which have the dimensions of curvature, it is easier to find
variables which are gauge-invariant in the background, and hence
natural candidates to perturb.   It is convenient nevertheless to
begin by perturbing the third-order Einstein-Ricci system (without
fixing the slicing term) and later to
take a time derivative to reach the perturbed fourth-order theory.
This procedure organizes and simplifies the computation because
the background is time-independent. 

As perturbations, let $g_{ij} = {\underline{g}}_{ij} + {g'}_{ij}$, $K_{ij}
= {\underline{K}}_{ij} + {K'}_{ij}$, ${\bar\Gamma}^i\mathstrut_{jk} =
{\underline{\bar\Gamma}}^i\mathstrut_{jk} +
{\bar\Gamma}^{\prime\,i}\mathstrut_{jk}$, where an underline indicates the
background quantity and a prime the first order perturbation. In the
fourth-order theory, the lapse is unperturbed, so $N\equiv
{\underline{N}}$, and the underline will be suppressed. The shift
$\beta^k\equiv {\underline{\beta}}^k$ is also unperturbed. 

To be explicit, consider perturbations of the Schwarzschild background
\begin{equation}
ds^2 = -(1 -{2 M\over r}) dt^2 + (1- {2M\over r})^{-1} dr^2 + r^2 d\theta^2
+ r^2 \sin^2 \theta\phi^2.
\end{equation}  
This example admits three useful simplifications: The background shift
vanishes, $\beta^k=0$, so ${\hat\partial_0} = {\partial}_t$. The background
lapse $N$ is time-independent, so many terms in (\ref{weq}) and
(\ref{weq2}) vanish. Finally, the background extrinsic curvature
${\underline{K}}_{ij}$ and its derivatives (space and time) vanish on the
natural slice ($t=$constant) to be perturbed. This implies that their Lie
derivatives along an arbitrary vector ${\bf v}$ vanish, {\it e.g.}
\begin{equation}
{\mathcal{L}}_{\bf v} {\underline{K}}_{ij} = 
v^k {\partial}_k {\underline{K}}_{ij} 
+ {\underline{K}}_{kj} {\partial}_i v^k 
+ {\underline{K}}_{ik} {\partial}_j v^k
=0.
\end{equation}
Thus, they are perturbatively gauge-invariant variables, and 
therefore natural 
candidates for perturbation.  Their selection accords
with the general principle emphasized in \cite{ElB} that one should
always choose to perturb quantities which vanish in the background.
In addition, many more terms in (\ref{weq}) and (\ref{weq2}) vanish 
because the extrinsic curvature is zero in the background.

The result of perturbing (\ref{weq}) in a background sharing
these properties (time-inde\-pendent lapse, vanishing shift, and
vanishing background extrinsic curvature) with Schwarzschild is 
the wave equation
\begin{eqnarray}
\label{pweq}
N \Omega'_{ij} &=& -{\partial}_t^2 {K'}_{ij} 
+ N^2 {\underline{\bar\nabla}}^k {\underline{\bar\nabla}}_k {K'}_{ij} 
-N^2 {\underline{\bar\nabla}}_i {\underline{\bar\nabla}}_j {H'} 
- 4 N {\underline{\bar\nabla}}^k N {\underline{\bar\nabla}}_{(i} {K'}_{j)k} \\
&&+ 3 N ({\underline{\bar\nabla}}^k N) {\underline{\bar\nabla}}_k {K'}_{ij} 
-2 N {K'}_{k(i} {\underline{\bar\nabla}}_{j)} {\underline{\bar\nabla}}^k N
-2 {K'}_{k(i} ({\underline{\bar\nabla}}_{j)} N) 
 {\underline{\bar\nabla}}^k N \nonumber \\
&&+ N {H'} {\underline{\bar\nabla}}_i {\underline{\bar\nabla}}_j N
+ {K'}_{ij} ({\underline{\bar\nabla}}_k N) {\underline{\bar\nabla}}^k N   
-2 N^2 {K'}_{k(i} {\underline{\bar R}}_{j)}\mathstrut^k 
-2 N^2 {\underline{\bar R}}^k\mathstrut_{ij}\mathstrut^m {K'}_{km} .
\nonumber
\end{eqnarray}
(An additional term ${K'}_{ij} N {\underline{\bar\nabla}}^k
{\underline{\bar\nabla}}_k N$ vanishes in the Schwarzschild background and
is not included.) The spatial Ricci and Riemann tensors are computed in the
background metric.  

Note the presence of the slicing term as the second derivative of ${H'}
={\mathrm{tr}}{{\mathbf {K'}}}$. This spoils the hyperbolicity of the
perturbative wave equation. If one wished, one could return to the full
theory and impose harmonic slicing. When perturbing the theory, one would
have to allow perturbations of the lapse, but this second derivative term
would be removed. Interestingly, the perturbed lapse would still not appear
in the perturbed wave equation. If one went on to introduce a tensor
spherical harmonic multipole decomposition of the perturbations of the
extrinsic curvature, as will be done below for the time-derivative of the
extrinsic curvature, perturbed third-order equations would result. These
form the basis of the perturbative-matching outer-boundary module being
used as one method of extracting gravitational waves and imposing outer
boundary conditions on the three-dimensional numerical simulations carried
out by the Binary Black Hole Alliance\cite{bbh97}. 

The second derivative of $H$ in the full theory can also be removed 
by passing to the fourth-order theory by taking a time-derivative and adding
second spatial derivatives of $R_{00}$.  Because the background is
time-independent and the background extrinsic curvature vanishes, 
the effect of taking a time-derivative is simply
to replace the perturbed extrinsic curvature ${K'}_{ij}$ by
its time derivative ${\partial}_t {K'}_{ij} = {\dot K'}_{ij}$.  The
additional perturbed $R_{00}$ term has the form
\begin{eqnarray}
N {\underline{\bar\nabla}}_i {\underline{\bar\nabla}}_j {R'}_{00}
 &=& N^2 {\underline{\bar\nabla}}_i {\underline{\bar\nabla}}_j {\dot H'} 
 + N {\dot H'} {\underline{\bar\nabla}}_i {\underline{\bar\nabla}}_j N 
+ 2 N ({\underline{\bar\nabla}}_{(i} N) 
   {\underline{\bar\nabla}}_{j)} {\dot H'} \\
&&\quad\quad
 - N {\underline{\bar\nabla}}_i 
 {\underline{\bar\nabla}}_j (N {\underline{g}}^{mn} {g'}_{nj} 
 {\underline{g}}^{jk} {\underline{\bar\nabla}}_m {\underline{\bar\nabla}}_k N
+N  {\underline{g}}^{mk} {\bar\Gamma}^{\prime\,l}\mathstrut_{mk} 
  {\partial}_l N) 
\nonumber 
\end{eqnarray}

One sees that by using ${R'}_{00}$, the second derivative of ${\dot H'}$ is
cancelled, but as well metric perturbations are introduced into the
perturbed fourth-order equations through ${g'}$ and $\bar\Gamma'$.   
This is the price that one has paid to achieve explicit hyperbolicity 
in the full theory. The next step is to eliminate the metric perturbations
in favor of time-differentiated extrinsic curvature perturbations.
This is done by using ${R'}_{00}$.  On first sight, this appears
simply to undo the step which cancelled the second derivative of
$H$, but this is not so.  We desire to understand the
behavior of the theory away from the constraint hypersurface, so
we do not set ${R'}_{00}=0$, but leave it as a free
variable which happens to vanish on-shell.  The presence of this
off-shell term marks the fact that the second derivative of $H$ has
been cancelled in the full theory. 

This raises a valuable point.
It is important to emphasize that one starts from a well-posed 
hyperbolic theory before perturbing.  If the
$R_{00}$ term were not present in the full theory, as it is not
when one simply takes a time-derivative of the third order
theory (without fixing the slicing), one would have a system which is 
believed not to be hyperbolic. 
The perturbed form of that non-hyperbolic system would
naively agree with the one we have just found if one 
were to set ${R'}_{00}=0$. 
This perturbative theory therefore agrees with general relativity on-shell,
but the Hamiltonian constraint has not been fully incorporated, so that
when violations of the Hamiltonian constraint occur, the theory 
cannot respond properly, that is, in a manifestly hyperbolic fashion.  

The $R_{00}$ term compensates behavior
of the second spatial derivative of $H$ to make the theory hyperbolic.
When $R_{00}$ vanishes, this compensation is evidently insignificant.
As the magnitude of the second spatial derivatives of $R_{00}$ increase 
however, one suspects that the compensation becomes more important.  
Because this condition involves second derivatives, 
small variations in $R_{00}$ from zero can nevertheless produce arbitrarily
large contributions to the equation.  This reveals an important
caveat:  it is dangerous
to study a constrained theory solely within the constraint hypersurface
and not to consider its behavior when the constraints are violated.
The full mathematical character of a fundamental physical theory
with constraints, in particular its hyperbolicity and well-posedness,
involves off-shell information.

Having given this warning, in the case of perturbations of
Schwarzschild at least, the role of the $R_{00}$ term appears to be 
relatively innocuous. 
The system of equations does not change character dramatically when this
term is removed.  This suggests there is further room to explore the
role of the $R_{00}$ term.

The equation for the perturbed fourth-order theory is 
\begin{eqnarray}
\label{pweq2}
N\tilde \Omega'_{ij} &=& -{\partial}_t^2 {\dot K'}_{ij} 
+ N^2 {\underline{\bar\nabla}}^k {\underline{\bar\nabla}}_k {\dot K'}_{ij}
-N^2 {\underline{\bar\nabla}}_i {\underline{\bar\nabla}}_j {\dot H'} 
+ N {\underline{\bar\nabla}}_i {\underline{\bar\nabla}}_j {R'}_{00}  \\
&& - 4 N {\underline{\bar\nabla}}^k N 
   {\underline{\bar\nabla}}_{(i} {\dot K'}_{j)k}
+ 3 N {\underline{\bar\nabla}}^k N {\underline{\bar\nabla}}_k {\dot K'}_{ij}  
-2 N {\dot K'}_{k(i} {\underline{\bar\nabla}}_{j)} 
   {\underline{\bar\nabla}}^k N
+ N {\dot H'} {\underline{\bar\nabla}}_i 
   {\underline{\bar\nabla}}_j N  \nonumber \\
&& -2 {\dot K'}_{k(i} ({\underline{\bar\nabla}}_{j)} N) 
    {\underline{\bar\nabla}}^k N
+ {\dot K'}_{ij} ({\underline{\bar\nabla}}_k N) {\underline{\bar\nabla}}^k N   
-2 N^2 {\dot K'}_{k(i} {\underline{\bar R}}_{j)}\mathstrut^k
-2 N^2 {\underline{\bar R}}^k\mathstrut_{ij}\mathstrut^m {\dot K'}_{km}  .
\nonumber
\end{eqnarray}
(Again, an additional term ${\dot K'}_{ij} N
{\underline{\bar\nabla}}^k{\underline{\bar\nabla}}_k N$ vanishes in
Schwarzschild case and has been dropped.)

\section{Odd-parity perturbations}

Because the wave operator on the Schwarzschild background separates
in spherical coordinates, perturbations in the Schwarzshild 
background are naturally handled by making a multipole decomposition
in tensor spherical harmonics.  These in turn naturally divide into
odd and even parity.  Following Moncrief\cite{Mon74}, we decompose 
the odd-parity perturbations as
\begin{equation}
{\dot K'}_{ij} = u_1(t,r) (\hat e_1)_{ij} + u_2(t,r) (\hat e_2)_{ij},
\end{equation} 
where $u_1(t,r)$ and $u_2(t,r)$ are radial perturbations and
\begin{equation}
\hat e_1 = \left( \begin{array}{ccc}
0&{\displaystyle -1\over \displaystyle \sin\theta}{\partial}_\phi {Y_{\ell m}}
&\sin\theta {\partial}_\theta {Y_{\ell m}} \\[0.2cm]
{\mathrm{symm}}&0&0 \\[0.2cm]
{\mathrm{symm}}&0&0
\end{array} \right)
\end{equation}
and
\begin{equation}
\hat e_2 =  \left( \begin{array}{ccc}
0&0&0 \\[0.3cm]
0&{\displaystyle 1\over \displaystyle \sin\theta}
({\partial}_\theta {\partial}_\phi - \cot\theta {\partial}_\phi) {Y_{\ell m}} & 
{\displaystyle \sin\theta\over \displaystyle 2} 
({\displaystyle 1\over \displaystyle \sin^2\theta} 
{\partial}_\phi^2 + \cot\theta {\partial}_\theta
-{\partial}_\theta^2) {Y_{\ell m}} \\[0.3cm]
0&{\mathrm{symm}}&-\sin\theta 
  ({\partial}_\theta {\partial}_\phi - \cot\theta {\partial}_\phi) {Y_{\ell m}}
\end{array} \right)
\end{equation}
are the odd-parity tensor spherical harmonics.  In these formulae,
${Y_{\ell m}}(\theta,\phi)$ are the standard scalar spherical harmonics 
satisfying
$$ ({\partial}_\theta^2 + \cot\theta {\partial}_\theta 
+{1\over \sin^2 \theta} {\partial}_\phi^2) {Y_{\ell m}} =
-\ell(\ell+1) {Y_{\ell m}}.$$
The notation ``symm'' indicates the matrices are symmetric.

{}From (\ref{pweq2}) one obtains equations for $u_1$ and $u_2$.  To simplify
the expressions it is useful to introduce 
$${\lambda_\ell}=\ell(\ell+1)$$ 
as the (negative of the) eigenvalue of the spherical harmonics and to 
replace second and higher derivatives of $N$ by equivalent expressions
in terms of $N$ and $N'\equiv {\partial}_r N$ (not to confuse the prime with
the perturbative part).  From the $r\theta$-component of 
(\ref{pweq2}), after dividing out the common angular factor 
$-{\partial}_\phi {Y_{\ell m}}/\sin\theta$, one has
\begin{eqnarray}
\label{u1eq}
-{\partial}_t^2 u_1 + N^4 {\partial}_r^2 u_1 +  4 N^3 N' {\partial}_r u_1 + 
{N^2\over r^2} ( -{\lambda_\ell} - 4 N^2 + 6 r N N' + r^2 N^{\prime\,2})u_1 
&& \\
+ {(2- {\lambda_\ell}) N^2\over r^3} u_2 &=& 0 . \nonumber
\end{eqnarray}
{}From the $\theta\phi$ equation, after dividing out 
$$(-\sin\theta {\partial}_\theta^2 {Y_{\ell m}} 
+ \cos\theta {\partial}_\theta {Y_{\ell m}} 
+ \csc\theta {\partial}_\phi^2 {Y_{\ell m}})/2,$$
one has
\begin{eqnarray}
\label{u2eq}
-{\partial}_t^2 u_2 + N^4 {\partial}_r^2 u_2 
+{2 N^3\over r}(-N + 2 r N') {\partial}_r u_2 
\hspace{2cm}  && \\
+ {N^2 \over r^2}(-{\lambda_\ell} + 4 N^2 - 4 r N N' + r^2 N^{\prime\,2})u_2 
+ {4 N^3\over r}(-N + r N') u_1 &=& 0. \nonumber
\end{eqnarray}

Of the constraints, only one is present in odd parity.  It can
be found from $\dot {R'}_{0\theta}$, and its radial expression is
\begin{equation}
\label{cu}
{ \sin\theta \dot {R'}_{0\theta}\over {\partial}_\phi {Y_{\ell m}}} 
= c_{u\theta} =
N^3 {\partial}_r u_1 + ({2 N^3 \over r} + N^2 N') u_1 +
{{\lambda_\ell}-2\over 2 r^2 } N u_2 .
\end{equation}
(A degenerate expression is found from $\dot {R'}_{0\phi}$.)
If $c_{u\theta}=0$, the odd-parity part of the momentum constraints hold; if
not, $c_{u\theta}$ measures the violation.  The variable $u_2$ can be
eliminated from (\ref{u1eq}) using this constraint, and one
has
\begin{equation}
-{\partial}_t^2 u_1 + N^4 {\partial}_r^2 u_1 
+ (4 N^3 N' + {2 N^4\over r}) {\partial}_r u_1 + 
{N^2\over r^2} (-{\lambda_\ell} + 8 r N N' + r^2 N^{\prime\,2}) u_1 -
{2 N \over r} c_{u\theta} = 0. 
\end{equation}
By rescaling $u_1$, the differential part of the operator can be
brought into the familiar Regge-Wheeler form,   
$$-{\partial}_t^2 + N^2 {\partial}_r N^2 {\partial}_r.$$ 
The new variable
\begin{equation}
A_u \equiv r N u_1
\end{equation}
is the gauge-invariant variable which satisfies the Regge-Wheeler equation
\begin{equation}
\label{Au_eqn1}
\biggl( -{\partial}_t^2 + N^2 {\partial}_r N^2 {\partial}_r 
- {N^2 \over r^2}({\lambda_\ell} - 6 r N N') 
\biggr) A_u
-2 N^2 c_{u\theta} = 0.
\end{equation}
The familiar Regge-Wheeler potential is explicitly given by
\begin{equation}
\label{RWpot2}
V_{RW}(r) = {N^2 \over r^2}({\lambda_\ell} - 6 r N N') = 
(1 - {2 M\over r}) ({\ell(\ell+1)\over r^2}-{6M\over r^3}).
\end{equation}
The $c_{u\theta}$ term reflects how the Regge-Wheeler equation is modified when
the odd-parity momentum constraint is violated.

{}From the perturbative form of (\ref{curv}) in the Schwarzschild background,
the gauge-invariant variable $A_u$ is readily identified as the
radial part of one of the components of the Riemann tensor 
\begin{equation}
A_u = r N u_1 =- r {\sin\theta {R'}_{0r0\theta} \over  
{\partial}_\phi {Y_{\ell m}}}
\end{equation}
The ``extra'' factor of $r$ is an artifact which arises because the
differential part of the Regge-Wheeler operator is written as if
space were one-dimensional whereas the operator actually comes from the
radial part of a three-dimensional Laplacian.  The factor of $r$ is
just the scaling to change the apparent dimension of the background space.

The equation which evolves $c_{u\theta}$ is found 
by forming the differential linear combination of
(\ref{u1eq}) and (\ref{u2eq}) implied by (\ref{cu}) to create a second time 
derivative of $c_{u\theta}$.  One finds
\begin{equation}
\label{momu1}
\biggl( -{\partial}_t^2 + N^2 {\partial}_r N^2 {\partial}_r 
- {N^2 {\lambda_\ell}\over r^2} \biggr) c_{u\theta} = 0.
\end{equation}
(The scaling of $c_{u\theta}$ was chosen earlier to put the differential
part of the operator in Regge-Wheeler form.)

The odd-parity violation of the momentum constraints propagates
hyperbolically on its own, and the pair (\ref{Au_eqn1}) and (\ref{momu1})
together form a hyperbolic system. If $c_{u\theta}=0$ and ${\partial}_t
c_{u\theta}=0$ initially, (\ref{momu1}) guarantees that they remain so. It
is a necessary feature that the constraints evolve amongst themselves
without involving the physical degree of freedom. This enables zero initial
data to remain zero. We will see this again in the even-parity case.

\section{Even-parity perturbations}
\label{sec:eqns}

Following Moncrief, but relabelling the radial variables and rescaling
the $g_1$ variable to have more rational dimensions, we can decompose the 
even-parity perturbations as 
\begin{equation}
{\dot K'}_{ij} = r g_1(t,r) (\hat f_1)_{ij} + g_2(t,r) N^{-2} (\hat f_2)_{ij}
+ r^2 g_3(t,r) (\hat f_3)_{ij} + r^2 g_4 (\hat f_4)_{ij},
\end{equation}
where 
\begin{equation}
\hat f_1 = \left( \begin{array}{ccc}
0 & {\partial}_\theta {Y_{\ell m}} & {\partial}_\phi {Y_{\ell m}} \\[0.3cm]
{\mathrm{symm}} & 0 & 0 \\[0.3cm]
{\mathrm{symm}} & 0 & 0
\end{array} \right),
\end{equation}
\begin{equation}
\hat f_2 = \left( \begin{array}{ccc}
{Y_{\ell m}} & 0 & 0 \\[0.3cm]
0 & 0 & 0 \\[0.3cm]
0 & 0 & 0
\end{array} \right),
\end{equation}
\begin{equation}
\hat f_3 = \left( \begin{array}{ccc}
0 & 0 & 0 \\[0.3cm]
0 & {Y_{\ell m}} & 0 \\[0.3cm]
0 & 0 & \sin^2\theta\, {Y_{\ell m}} 
\end{array} \right),
\end{equation}
\begin{equation}
\hat f_4 = \left( \begin{array}{ccc}
0 & 0 & 0 \\[0.3cm]
0 & {\partial}_\theta^2 {Y_{\ell m}} &
({\partial}_\theta {\partial}_\phi - \cot\theta {\partial}_\phi) 
   {Y_{\ell m}}\\[0.3cm]
0 & {\mathrm{symm}} & 
({\partial}_\phi^2 + \sin\theta \cos\theta {\partial}_\theta) {Y_{\ell m}} 
\end{array} \right)
\end{equation}
are the even-parity tensor spherical harmonics.
The trace of ${\dot K'}_{ij}$ is found to be
\begin{equation}
{\dot H'} = (g_2 + 2 g_3 - {\lambda_\ell} g_4) {Y_{\ell m}}.
\end{equation}
Finally, let
$${R'}_{00} = c_h {Y_{\ell m}}. $$

The fourth-order Einstein-Ricci equations (\ref{pweq2}) for even-parity
radial perturbations have the following forms.
{}From the $r\theta$-equation, up to a 
factor of $r {\partial}_\theta {Y_{\ell m}}$, the equation for $g_1$ is
\begin{eqnarray}
\label{g1eq}
{\hat L}_1:&& -{\partial}_t^2 g_1 + N^4 {\partial}_r^2 g_1 + 
({2 N^4 \over r} + 4 N^3 N'){\partial}_r g_1  
 - {N^2 \over r}{\partial}_r (g_2 + 2 g_3 - {\lambda_\ell} g_4) \nonumber \\
&& \hspace{0.5cm} + {N^2 \over r^2}(-{\lambda_\ell} - 4 N^2 + 10 r N N' 
+  r^2 N^{\prime\,2})g_1  \\
&& + {N\over r^2}(3 N - 2 r N') g_2 +{N^2\over r^2}({\lambda_\ell}-2) g_4 
+ {N\over r} {\partial}_r c_h -{N\over r^2} c_h =0.
\nonumber
\end{eqnarray}
{}From the $rr$-equation, up to a factor of ${Y_{\ell m}}/N^2$, the equation
for $g_2$ is
\begin{eqnarray}
\label{g2eq}
{\hat L}_2:&&-{\partial}_t^2 g_2 
- N^4 {\partial}_r^2 (2 g_3 - {\lambda_\ell} g_4)  
+ ({2 N^4\over r} - N^3 N'){\partial}_r g_2 
- N^3 N' {\partial}_r (2 g_3 - {\lambda_\ell} g_4)  \nonumber \\
&&\hspace{0.5cm} +{4 N^4 {\lambda_\ell} \over r^2} g_1 
+ {N^2 \over r^2} (-{\lambda_\ell} -4 N^2 + 6 r N N' -
r^2 N^{\prime\,2}) g_2  \\
&& +{2N^3 \over r^2} ( N -2 r  N') (2 g_3 -{\lambda_\ell} g_4)
 + N^3 {\partial}_r^2 c_h + N^2 N' {\partial}_r c_h  = 0. \nonumber
\end{eqnarray}
{}From the $\theta\phi$-equation, up to a factor of $r^2 ({\partial}_\theta
{\partial}_\phi {Y_{\ell m}} -\cot\theta {\partial}_\phi {Y_{\ell m}})$,
the equation for $g_4$ is
\begin{eqnarray}
\label{g4eq}
{\hat L}_4:&& -{\partial}_t^2 g_4 + N^4 {\partial}_r^2 g_4 
+{2 N^3\over r}(N + 2 r N'){\partial}_r g_4 
+{4 N^3\over r^2}(N-r N')g_1 -{N^2\over r^2} g_2 - {2 N^2\over r^2} g_3 
\nonumber \\
&& \hspace{1cm}+ {N^2\over r^2}(2 N^2 + 4 r N N' +  r^2 N^{\prime\, 2} ) g_4
+ {N\over r^2} c_h
= 0. 
\end{eqnarray}
Finally, taking the trace of the $\theta\theta$- and $\phi\phi$-equations,
dividing out the common factor $2 {Y_{\ell m}}$, and adding
${\lambda_\ell}/2$ times (\ref{g4eq}), one obtains the equation for $g_3$ 
\begin{eqnarray}
\label{g3eq}
{\hat L}_3:&& -{\partial}_t^2 g_3 
+ N^4 {\partial}_r^2 g_3 -{N^4 \over r} {\partial}_r g_2 
+ 4 N^3 N' {\partial}_r g_3
+{{\lambda_\ell} N^4\over r} {\partial}_r g_4  \\
&&\hspace{0.5cm} + {N^3\over r^2}(2 N - 5 r N') g_2 
+ {N^2\over r^2}(-{\lambda_\ell}  - 2 N^2 + 6 r N N' +  r^2 N^{\prime\, 2}) g_3  
\nonumber \\
&& \hspace{0.5cm}  
+{{\lambda_\ell} N^3 \over r^2}(2 N -r N') g_4 
+ {N^3 \over r} {\partial}_r c_h = 0. \nonumber
\end{eqnarray}

The (time-differentiated) perturbed momentum constraints in 
even-parity reduce to the two expressions
\begin{equation}
\label{momR}
{\dot R'_{0r}\over N {Y_{\ell m}}} \equiv {\tilde c}_r 
= {\partial}_r (2 g_3 - {\lambda_\ell} g_4) 
+{1\over r}({\lambda_\ell} g_1 - 2 g_2 +2 g_3 - {\lambda_\ell} g_4),
\end{equation}
\begin{equation}
\label{momTh}
{\dot R'_{0\theta}\over N {\partial}_\theta {Y_{\ell m}}} 
\equiv {\tilde c}_{g\theta} = -r N^2 {\partial}_r g_1
-g_1 (3 N^2 + r N N') + g_2 + g_3 - g_4.
\end{equation}
(Again, a degenerate expression is found for $\dot R'_{0\phi}$.)
These define variables ${\tilde c}_r$ and ${\tilde c}_{g\theta}$ which are
useful for studying the effects of momentum constraint violations in
perturbation theory. By forming the appropriate differential linear
combinations of (\ref{g1eq})-(\ref{g3eq}), one finds that the momentum
constraint variables satisfy the following system of coupled wave
equations
\begin{eqnarray}
-{\partial}_t^2 {\tilde c}_r + N^4 {\partial}_r^2 {\tilde c}_r 
+ {2 N^3\over r }(N + 4 r N') {\partial}_r {\tilde c}_r  
+ {4 N^2 N'\over r} {\partial}_r c_h&& \\
+ {N^2 \over r^2 } (-2-{\lambda_\ell} +9 r^2 N^{\prime\, 2} ) {\tilde c}_r 
+ { 2 {\lambda_\ell} N^2 \over r^3 }  {\tilde c}_{g\theta}
- {{\lambda_\ell} N' \over r^2} c_h &=& 0, \nonumber 
\end{eqnarray}
\begin{eqnarray}
-{\partial}_t^2 {\tilde c}_{g\theta} 
+ N^4 {\partial}_r^2 {\tilde c}_{g\theta} 
+ 4 N^3 N' {\partial}_r {\tilde c}_{g\theta} 
- N^2 N' {\partial}_r c_h\hspace{2cm}&& \\
+ {N^2 \over r^2} (-{\lambda_\ell} - 2 r N N' 
+ r^2 N^{\prime\, 2} ) {\tilde c}_{g\theta} 
+ {2 N^3 \over r } ( N + r N' ) {\tilde c}_r
 &=& 0. \nonumber
\end{eqnarray}
Introducing rescaled momentum constraint variables 
\begin{equation}
c_r = r N^3 {\tilde c}_r = { r N^2 \dot R'_{0r}\over {Y_{\ell m}}}
\end{equation}
and 
\begin{equation}
c_{g\theta} = N {\tilde c}_{g\theta} 
= {\dot R'_{0\theta}\over {\partial}_\theta {Y_{\ell m}}}
\end{equation}
puts the differential operator part in Regge-Wheeler form, leaving
the equations
\begin{eqnarray}
\label{momr1}
\biggl(-{\partial}_t^2 + N^2 {\partial}_r N^2 {\partial}_r  
- {N^2 \over r^2}(2+{\lambda_\ell} + 2 r N N')
\biggr) c_r  \hspace{2cm}&& \\
+ {2 N^4 {\lambda_\ell}\over r^2} c_{g\theta} + 4 N^5 N' {\partial}_r c_h
- {{\lambda_\ell} N^3 N'\over r} c_h &=& 0
\nonumber
\end{eqnarray}
and
\begin{equation}
\label{momg1}
\biggl( -{\partial}_t^2 + N^2 {\partial}_r N^2 {\partial}_r 
-{{\lambda_\ell} N^2 \over r^2} \biggr) c_{g\theta} 
+ {2 N\over r^2}( N + r N') c_r 
- N^3 N' {\partial}_r c_h = 0.
\end{equation}

An evolution equation for the $R_{00}$ constraint is found by taking
the trace of (\ref{pweq2}).  The perturbed constraint is 
\begin{equation}
\label{chdef}
c_h {Y_{\ell m}} = {R'}_{00} = N {\dot H'} 
-N {\underline{g}}^{mn} {g'}_{nj} {\underline{g}}^{jk} 
  {\underline{\bar\nabla}}_m {\underline{\bar\nabla}}_k N
- N {\underline{g}}^{mk} {\bar\Gamma}^{\prime\,i}\mathstrut_{mk} 
  {\underline{\bar\nabla}}_i N.
\end{equation} 
This involves metric perturbations through the ${g'}$ and ${\bar\Gamma'}$
terms. A time derivative of ${\bar\Gamma'}$ only involves perturbed
extrinsic curvatures and a second time derivative involves ${\dot
K'}_{ij}$,
\begin{equation}
{\partial}_t^2 {\bar\Gamma}^{\prime\,i}\mathstrut_{mk} = 
-{\underline{g}}^{ij} \biggl({\underline{\bar\nabla}}_m (N{\dot K'}_{jk}) 
+ {\underline{\bar\nabla}}_k (N{\dot K'}_{jm}) -
{\underline{\bar\nabla}}_j (N{\dot K'}_{mk})\biggr).
\end{equation}
Taking two time derivatives of (\ref{chdef}) allows second time derivatives
of ${\dot H'}$ to be replaced by second derivatives of $c_h$ and leads to
the equation
\begin{equation}
\label{hameq1}
\biggl(-{\partial}_t^2  + N^4 {\partial}_r^2 
+ ({2N^4\over r} + N^3 N'){\partial}_r
-{ N^2 {\lambda_\ell}\over r^2} \biggr) c_h + {2 N N'\over r} c_r=0.
\end{equation}

An alternative derivation which reveals the role of the Hamiltonian
constraint $G^0\mathstrut_0$ is given as follows.  Let 
\begin{equation}
\label{chGdef}
c_{hG} {Y_{\ell m}} = {G'}^0\mathstrut_0 = - {1\over 2} {\bar R'}.
\end{equation}
Two time derivatives gives
\begin{equation}
{\partial}_t^2 (c_{hG} {Y_{\ell m}}) 
= - {\underline{\bar\nabla}}^k \dot {R'}_{k0} 
- N^{-1} ({\underline{\bar\nabla}}^k N)\dot {R'}_{k0},
\end{equation}
which is simply the time derivative of the contracted Bianchi identity
$\nabla^\mu G_{\mu 0}$ as mentioned above.  The trace of (\ref{pweq2})
corresponds through (\ref{weq2}) to
\begin{equation}
\label{trpweq2}
{\partial}_t^2 {R'}^k\mathstrut_k - 2 {\underline{\bar\nabla}}^k \dot {R'}_{k0} 
+ {\underline{\bar\nabla}}^k {\underline{\bar\nabla}}_k R_{00}
=0.
\end{equation}
The second time derivative of the perturbed form of (\ref{GRid})
gives
\begin{equation}
{\partial}_t^2 {R'}^k\mathstrut_k = -N^{-2} {\partial}_t^2 (c_h {Y_{\ell m}}) 
- 2 {\partial}_t^2 (c_{hG} {Y_{\ell m}}).
\end{equation}
Substituting this and (\ref{chGdef}) into (\ref{trpweq2}) gives 
\begin{equation}
\label{hameq2}
-N^{-2} {\partial}_t^2 (c_h {Y_{\ell m}}) 
+ {\underline{\bar\nabla}}^k {\underline{\bar\nabla}}_k (c_h {Y_{\ell m}}) 
+ 2 N^{-1} ({\underline{\bar\nabla}}^k N) \dot {R'}_{k0}
=0,
\end{equation}
which is readily confirmed to agree with (\ref{hameq1}).

The constraint variables $c_r$, $c_{g\theta}$, and $c_h$ together form
a hyperbolic system.  If they and their first time derivatives all
vanish, the initial data for this system is identically zero and
it remains zero under evolution.   
If any are non-zero, the constraint violations evolve hyperbolically.
Below we will see that $c_r$, $c_{g\theta}$ and $c_h$ occur as source
terms in the even-parity Regge-Wheeler equation. 

\section{Intertwining and the even-parity Regge-Wheeler equation}

Having found how constraint violations evolve, it remains to
find how the physical even-parity degree of freedom
propagates.  The procedure is to form an arbitrary first-order differential
linear combination of (\ref{g1eq})-(\ref{g3eq}) and require
that it equal a Regge-Wheeler wave operator
\begin{equation}
\Box_{RW}  =-{\partial}_t^2 + N^2 {\partial}_r N^2 {\partial}_r - V_{RW}(r) 
\end{equation}
acting on
differential linear combination of the $g_k$, plus constraint variables. 
More precisely, one requires 
\begin{equation}
\label{ioRW}
\sum_{k=1}^4 \biggl(b_k(r) {\partial}_r + a_k(r) \biggr) {\hat L}_k =
\Box_{RW}  \biggl(\sum_{k=1}^4 b_k(r) {\partial}_r g_k + a_k(r) g_k \biggr) 
+ {\hat f}({\tilde c}_r,{\tilde c}_{g\theta},c_h).
\end{equation}
As discussed in the introduction, this equation can be understood as a
component of a matrix intertwining relation between a matrix differential
operator in the basis of the $g_k$ and one in the basis which isolates the
physical degree of freedom $A_g$ from the constraint variables ${\tilde
c}_r$, ${\tilde c}_{g\theta}$ and $c_h$. The other components of the matrix
transformation were determined above when the appropriate combinations of
the $g_k$ were found which produce equations involving only the constraint
variables. 

One can begin with the potential $V_{RW}(r)$ of the wave operator 
undetermined and solve for it self-consistently, but this complicates some of
the intermediate expressions, so the ansatz is made that
the potential has the Regge-Wheeler form (\ref{RWpot2}).  The ansatz for
the transformation (\ref{ioRW}) can be simplified slightly before
beginning.  Because the momentum constraint variables are built from
radial derivatives of $g_1$ and $2 g_3 - {\lambda_\ell} g_4$, 
one can without loss of generality set $b_1(r)=0$ and $b_3(r)=0$ 
because the momentum constraint variables can be used to eliminate
the radial derivatives of $g_1$ and $g_3$.   

The procedure is now to write out the equation (\ref{ioRW}) and compare
like derivatives. The collection of coefficients of the derivatives of the
$g_k$, starting from the highest derivatives, form a set of recursive
relations for the coefficients in the intertwining transformation. For
example, one immediately sees from the ${\partial}_r^3 g_2$ and
${\partial}_r^2 g_2$ coefficients that $b_2(r)=0$ and $a_2(r)=0$. Thus
$g_2$ does not appear in the physical even-parity gauge-invariant variable.
As each successive intertwining coefficient is determined, the remaining
relations are simplified. 

In anticipation of future developments, it proves convenient to rescale 
the remaining coefficients as follows
\begin{eqnarray}
b_4(r) &=& N^3 {\tilde b}_4(r),  \\
a_1(r) &=& N^3 {\tilde a}_1(r), \nonumber \\
a_3(r) &=& N {\tilde a}_3(r), \nonumber \\
a_4(r) &=& N {\tilde a}_4(r). \nonumber
\end{eqnarray}
It is also necessary to add zero in the form of an arbitrary multiple
$c_1(r),\ c_2(r)$ of the (unscaled) momentum constraint variables ${\tilde
c}_r, {\tilde c}_{g\theta}$ minus their expressions in terms of the $g_k$.
This reflects the fact that the right-hand side of (\ref{ioRW}) may contain
an arbitrary amount of momentum constraint variables. The Hamiltonian
constraint variable is not added as it would mix in metric perturbations
which are not present. [Note however that the Hamiltonian constraint
variable appears on both sides of (\ref{ioRW}) because of its presence in
(\ref{g1eq})-(\ref{g3eq}).]

After these simplifications, reduce second and higher derivatives of
$N$ to first and no derivatives by taking derivatives of $N^2=1-2M/r$, 
and denote differentiation of coefficients with respect to $r$ by primes. 
Then the set of recursive equations obtained from (\ref{ioRW}) becomes
\begin{eqnarray}
{2 N^7\over r}{\partial}_r^2 g_4:&\quad\quad& {\tilde b}_4 - r {\tilde b}'_4 \\
{N^5\over r^2} {\partial}_r g_2:&& r{\tilde a}_1 + r {\tilde a}_3 
+ {\tilde b}_4 \nonumber \\
{N^5\over r^2} {\partial}_r g_4:&& 
{\lambda_\ell} (r {\tilde a}_1 + r{\tilde a}_3 + {\tilde b}_4) 
+ 2 r {\tilde a}_4 - 2 r^2 {\tilde a}'_4 \nonumber \\
&&\quad + 4 r {\tilde b}_4 N N' - 8 r^2 {\tilde b}'_4 N N' 
- r^2 {\tilde b}''_4 N^2 
-{\lambda_\ell} r c_1 /2 \nonumber \\
-{2 N^5\over r^2} {\partial}_r g_3:&& 
 r {\tilde a}_1 + {\tilde b}_4 + r^2 {\tilde a}'_3 -r c_1/2 
\nonumber \\
{2 N^5\over r^2} {\partial}_r g_1:&& 
r {\tilde a}_1 N^2 + 2 {\tilde b}_4 N^2 - r^2 {\tilde a}'_1 N^2
-2 r^2 {\tilde a}_1 N N' - 2 r {\tilde b}_4 N N' + N r^3 c_2/2 \nonumber \\
{N^3\over r^3} g_2:&& 
-r {\tilde a}_4 + 3 r {\tilde a}_1 N^2 + 2r {\tilde a}_3 N^2 
+ 2 {\tilde b}_4 N^2
\nonumber \\
&&\quad - 2 r^2 {\tilde a}_1 N N' - 5 r^2 {\tilde a}_3 N N' 
-2 r {\tilde b}_4 N N' 
-r^3 c_2 N - r c_1 N^2\nonumber \\
{N^3 \over r^3} g_4:&& 
r {\lambda_\ell}(  {\tilde a}_4 +   {\tilde a}_1 N^2 + 2 {\tilde a}_3 N^2
-r {\tilde a}_3 N N' -c_1 N^2/2) \nonumber \\
&&\quad - 2  N^2 ( r {\tilde a}_1 - r {\tilde a}_4 
+2 {\tilde b}_4 N^2 + 2 r {\tilde b}_4 N N' -
2 r^2 {\tilde b}_4 N^{\prime\,2} ) \nonumber \\
&&\quad - 4 r^3 {\tilde a}'_4 N N'  - r^3 N^2 {\tilde a}''_4 
+ r^3 c_2 N \nonumber \\
{N^3 \over r^3} g_3:&&
-2 r {\tilde a}_4 - 2 r{\tilde a}_3 N^2 + 4 {\tilde b}_4 N^2 +
2 r^2 {\tilde a}_3 N N' - 4 r {\tilde b}_4 N N' \nonumber \\
&&\quad - 4 r^3 {\tilde a}'_3 N N' -r^3 {\tilde a}''_3 N^2 
+ r c_1 N^2 - r^3 c_2 N
\nonumber \\
{N^3\over r^3} g_1:&& 
4 N^2 (  r {\tilde a}_4 -  r {\tilde a}_1 N^2 - 2 {\tilde b}_4 N^2)
+ N^3 N' (10 r^2 {\tilde a}_1  + 28 r {\tilde b}_4 
- 8 r^3 {\tilde a}'_1) \nonumber \\
&&\quad - 8 r^2 N^2 N^{\prime\,2} ( r {\tilde a}_1  + {\tilde b}_4)
-4 r^2 N N' {\tilde a}_4 - r^3 {\tilde a}''_1 N^4 \nonumber \\
&&\quad + {\lambda_\ell} r c_1 N^2/2 + 3 r^3 c_2 N^3 
+ r^4 c_2 N^2 N'
\nonumber 
\end{eqnarray}

{}From the ${\partial}_r^2 g_4$ coefficient, one has the relation
\begin{equation}
{\tilde b}_4 = r {\partial}_r {\tilde b}_4
\end{equation}
from which one concludes ${\tilde b}_4 =r$.  (An overall multiplicative
constant can be
taken to be unity without loss of generality since it represents a
constant scale of the variable $A_g$.)  To cancel the ${\partial}_r g_2$
term, one must set
\begin{equation}
{\tilde a}_3(r) = -{\tilde a}_1(r) -1.
\end{equation}
Cancelling the ${\partial}_r g_3$ and ${\partial}_r g_1$ terms determines
$c_1$ and $c_2$ to be
\begin{eqnarray}
c_1(r) &=& 2 (1 +{\tilde a}_1 - r {\partial}_r {\tilde a}_1) \\
c_2(r) &=& {-2\over r^2} \biggl(2 N - 2 r N' + {\tilde a}_1 (N- 2 r N') 
-r N  {\tilde a}'_1 \biggr). \nonumber 
\end{eqnarray}
The coefficient of $g_2$ determines ${\tilde a}_4$ in terms of ${\tilde a}_1$
\begin{equation}
{\tilde a}_4 = 2 N^2 - r N N' + {\tilde a}_1 (N^2 -r N N').
\end{equation}

After imposing this, the coefficient of ${\partial}_r g_4$ gives the
equation
\begin{equation}
-{\lambda_\ell} + 4 N^2 - 16 r N N' 
+ {\tilde a}_1 (-{\lambda_\ell} + 2 N^2 - 8 r N N')
+ r  {\tilde a}'_1 ({\lambda_\ell} -2 N^2 + 2 r N N') = 0.
\end{equation}
Inspection suggests the substitution 
$${\tilde a}_1 = \alpha r -2$$
which leads to 
$$-6 M \alpha + {\lambda_\ell} =0.$$
Thus, one has
\begin{equation}
{\tilde a}_1 = {{\lambda_\ell} r\over 6 M} -2.
\end{equation}
After applying this relation, all the intertwining coefficients are
fixed.  The remaining equations implied by the $g_1$, $g_3$ and $g_4$
coefficients are satisfied identically if the lapse takes
the Schwarzschild form $N^2 = 1 - 2M/r$.  These final relations are
the consistency conditions of the over-determined system of equations.
It is easy to see that if the lapse weren't special, e.g. Schwarzschild,
the intertwining would not have been consistent. 

Assembling the above results gives the physical even-parity gauge-invariant
variable
\begin{equation}
A_g = N \biggl( r N^2 {\partial}_r g_4 
+ N^2 ({r{\lambda_\ell} \over 6 M} -2) g_1
+ (1 - {r {\lambda_\ell}\over 6 M}) g_3 
 + ({M\over r} - {{\lambda_\ell}\over 2} +
{r{\lambda_\ell}\over 6 M}) g_4 \biggr) . 
\end{equation}
This variable satisfies the Regge-Wheeler equation
\begin{eqnarray}
\label{Ag_eqn1}
\biggl( -{\partial}_t^2 
+ N^2 {\partial}_r N^2 {\partial}_r - V_{RW}(r)\biggr) A_g 
+ {N^2 \over r^2} c_r \hspace{3cm} && \\
+ {N^2\over r^2} ({2 {\lambda_\ell} \over 3 } - {4 M  \over r}) c_{g\theta}  
+ {N^2\over r^2} ({2 M\over r} - { {\lambda_\ell} \over 6 }) c_h 
&=& 0. \nonumber
\end{eqnarray}
As with the odd-parity equation, there are constraint violating
terms present to show how the theory evolves off-shell (perturbatively).
On-shell this is the Regge-Wheeler equation.  The gauge-invariant
variable $A_g$ has even-parity and is constructed from time-derivatives
of the extrinsic curvature.  

One can reach the Zerilli equation
\begin{equation}
\biggl( -{\partial}_t^2 
+ N^2 {\partial}_r N^2 {\partial}_r - V_{Z}(r)\biggr) \tilde A_g +
\cdots = 0,
\end{equation}
where $\tilde A_g = D_1 A_g$ and the Zerilli potential is
\begin{equation}
V_Z(r) = {2 N^2 \over r^3 ( n r + 3 M)^2} \biggl((n+1)n^2 r^3 + 3 M n^2 r^2 +
9  M^2 n r + 9 M^3\biggr),
\end{equation}
($n=(\ell-1)(\ell+2)/2$) by applying the appropriate intertwining 
operator $D_1$ (\ref{ioRWZer}) to (\ref{Ag_eqn1}) and 
using (\ref{RWtoZ}).  Note the Zerilli 
potential has an unphysical singularity at $nr + 3M=0$.
Making this final transformation is unnecessary and arguably undesirable
because of the unphysical singularity structure of the potential.  
Both the
odd- and even-parity gauge-invariant variables satisfy a Regge-Wheeler
equation, so parity is not the explanation for the existence
of the Zerilli equation.  Whether the existence of the Zerilli
equation has a deeper meaning than accidental coincidence is an open 
question. From the standpoint of computation, however, the physics
of even-parity perturbations is fully captured in the Regge-Wheeler
equation without spurious unphysical complications, and the
choice of description should be clear. 

\section{Conclusion}
 
For completeness, the full set of gauge-invariant perturbation equations 
(\ref{Au_eqn1}), (\ref{momu1}), (\ref{momr1}), (\ref{momg1}), (\ref{hameq1}), 
(\ref{Ag_eqn1}) is
gathered here
\begin{equation}
\label{Au_eqn}
\biggl(-{\partial}_t^2 + N^2 {\partial}_r N^2 {\partial}_r
- {N^2 \over r^2}({\lambda_\ell} - 6 r N N')\biggr) A_u
-2 N^2 c_{u\theta} = 0.
\end{equation}
\begin{eqnarray}
\label{Ag_eqn}
\biggl(-{\partial}_t^2 + N^2 {\partial}_r N^2 {\partial}_r
- {N^2 \over r^2}({\lambda_\ell} - 6 r N N') \biggr)
A_g \hspace{2cm}&& \\
 + {N^2 \over r^2} c_r
+ {N^2 \over r^2}({2 {\lambda_\ell}  \over 3 } - {4 M  \over r}) c_{g\theta}
+ {N^2\over r^2} ({2 M\over r} - { {\lambda_\ell} \over 6 }) c_h 
&=& 0. \nonumber
\end{eqnarray}
\begin{equation}
\label{momu}
\biggl(-{\partial}_t^2 
+ N^2 {\partial}_r N^2 {\partial}_r 
- {N^2 \over r^2}{\lambda_\ell} \biggr) c_{u\theta} = 0.
\end{equation}
\begin{equation}
\label{hameq}
\biggl(-{\partial}_t^2  
+ N^4 {\partial}_r^2 + ({2N^4\over r} + N^3 N'){\partial}_r
-{ N^2 {\lambda_\ell}\over r^2} \biggr) c_h + {2 N N'\over r} c_r=0.
\end{equation}
\begin{equation}
\label{momr}
\biggl(-{\partial}_t^2 
+ N^2 {\partial}_r N^2 {\partial}_r 
- {N^2 \over r^2}(2+{\lambda_\ell} + 2 r N N')
\biggr) c_r
+ {2 N^4 {\lambda_\ell}\over r^2} c_{g\theta}
+\biggl(  4 N^5 N' {\partial}_r - {{\lambda_\ell} N^3 N'\over r}\biggr) c_h = 0
\end{equation}
\begin{equation}
\label{momg}
\biggl(-{\partial}_t^2  
+ N^2 {\partial}_r N^2 {\partial}_r 
-{{\lambda_\ell} N^2 \over r^2}\biggr) c_{g\theta}
+ {2 N\over r^2}( N + r N') c_r
- N^3 N' {\partial}_r c_h
= 0.
\end{equation}

{\sloppy
Several important lessons are learned from the example of perturbing
the Schwarzschild black hole.  The one specific to the 
Schwarzschild
case is that the Regge-Wheeler equation suffices to describe the evolution of 
both odd and even parity physical perturbations.  This result has been known 
mathematically for some time\cite{Cha83,AnP,Whi}, but a skeptic might assign 
significance to the observation that the Zerilli equation always seems
to arise from even-parity perturbations\cite{Cha83,Mon74,Zer70}.  
The calculation presented here should answer that doubt. 
The Regge-Wheeler equation has been obtained directly
in even parity by an intertwining transformation, which decouples
the equations and involves the fewest number of derivatives possible.  
The intertwining procedure is an effective tool which systematizes
a technically involved computation and makes it transparent.
The Zerilli equation can be reached by a further 
transformation which introduces an unphysical angular momentum-dependent 
singularity into the potential.  Chandrasekhar\cite{Cha83} (pp. 198-199)
discusses whether one should dismiss the Zerilli equation and concludes
not.  ``Dismiss'' is probably too strong a word, but one certainly
need never use the Zerilli equation in computations, as all of the
physics in it is captured more succinctly in the Regge-Wheeler
equation. 
}

Perhaps the principal lesson which follows from the Schwarzschild example
is the recognition that working with the fourth-order Einstein-Ricci 
formulation shifts the emphasis from metric perturbations to 
curvature perturbations and in so doing reveals that the physical 
gauge-invariant quantities propagated by the Regge-Wheeler 
equation are formed from curvature perturbations and their spatial 
derivatives.  Furthermore, one discovers that the gauge-invariant 
perturbations in unphysical directions are given by (time derivatives
of) violations of the linearized Hamiltonian and momentum constraints.  
Additionally, the constraint violations evolve as a closed hyperbolic 
subsystem (\ref{momu})-(\ref{momg}), so that if the constraints are 
satisfied initially, they continue to be satisfied.
It is worth emphasizing that this evolution subsystem for the 
constraints comes directly from the dynamical equations 
and not from separately considering the contracted Bianchi identities.  
The information
for evolving the constraints is embedded within the dynamical part of
the theory, as it must be for consistency.

These observations are general and help one to organize the calculation of 
gauge-invariant perturbations in a wide class of constrained hyperbolic 
theories:  After recasting a constrained theory in hyperbolic form 
without gauge-fixing, say by the procedure outlined in \cite{aacby95b}, 
one perturbs a complete set of independent combinations of the 
fundamental variables which vanish in the background.  Since
the combinations vanish in the background, their perturbations
are necessarily gauge-invariant\cite{ElB}.  The constraints 
themselves provide a subset of these variables, and their
perturbations will evolve as a closed subsystem if 
the equations are consistent.  From a complete set of gauge-invariant
variables, the physical subset which are independent of the constraint
variables can be constructed in principle by using the intertwining procedure
described in the text. 

{\sloppy
The fact that the gauge-invariant perturbations of Minkowski space
and the Schwarzschild black hole
are curvature perturbations which split into the true linearized
degrees of freedom and constraints encourages speculation
about the nature of the true degrees of freedom in the fully nonlinear
theory.  From the twenty degrees of freedom of the Riemann tensor
in four dimensions, the ten components of the Ricci tensor are
obtained by tracing with the spatial 3-metric ($R^0\mathstrut_{i0j}$
must of course be added to $R^k\mathstrut_{ikj}$ to obtain
the spatial Ricci tensor).  Respectively, this accounts for one, three and 
all six degrees of freedom of $R_{0i0j}$, $R_{0ijk}$ and $R_{ijkl}$.
The ten Ricci components vanish (in vacuum) by equivalence 
of the Einstein-Ricci formulation with Einstein's theory. Six 
more degrees of freedom are obtained, three each, from divergences of 
$R_{0i0j}$ and $R_{0ijk}$. By the Bianchi identities, these reduce to 
combinations of derivatives of the Ricci tensor and hence also vanish.
This leaves two degrees of freedom each in $R_{0i0j}$ and $R_{0ijk}$,
namely the transverse traceless parts.  Work is in progress to prove the 
conjecture that something related to the transverse traceless parts of 
the Riemann tensor are the true degrees of freedom of the gravitational
field\cite{AnY}. 
}
\vspace{1cm}

Acknowledgment.  We would like to thank our close collaborators
J.W. York and Y. Choquet-Bruhat for many important discussions and
C.R. Evans and M. Rupright for contributions to the early stages
of this work.   This work was supported in part by 
the National Science Foundation,  grants PHY 94-13207 and 
PHY 93-18152/ASC 93-18152 (ARPA supplemented).

\end{document}